\documentclass[prb,aps,twocolumn]{revtex4-1}
\usepackage{amssymb,bm}
\usepackage[dvips]{graphicx}
\usepackage[dvips]{color}
\usepackage{amsmath} 
\usepackage{setspace}
\usepackage{enumerate}
\usepackage[none]{hyphenat}

\begin{document}

\title{Is Silicene the Next Graphene?}

\author{L. C. Lew Yan Voon$^1$ and G. G. Guzm\'{a}n-Verri$^2$}
\affiliation{$^1$School of Science and Mathematics, The Citadel, Charleston, SC 29409\\
$^2$Materials Science Division, Argonne National Laboratory, Argonne, IL 60439}

\begin{abstract}
This article reviews silicene, a relatively new allotrope of silicon, which can also be viewed
as the silicon version of graphene. Graphene is a two-dimensional material with unique
electronic properties qualitatively different from those of standard semiconductors such as
silicon. While many other two-dimensional materials are now being studied, our focus here
is solely on silicene. We first discuss its synthesis and the challenges presented. Next, a
survey of some of its physical properties is provided. Silicene shares many of the fascinating
properties of graphene, such as the so-called Dirac electronic dispersion. The slightly different
structure, however, leads to a few major differences compared to graphene, such as the
ability to open a bandgap in the presence of an electric field or on a substrate, a key property
for digital electronics applications. We conclude with a brief survey of some of the potential
applications of silicene. 
\end{abstract}

\date{\today}
\maketitle

\section{Introduction}

The study of nanomaterials took off some 40 years ago with
the design of so-called quasi-two-dimensional (2D) (solids
thin in one direction, typically around tens of nanometers
or tens of atomic layers in thickness), quasi-one-dimensional
(nanowires), and quasi-zero-dimensional solids (quantum dots).
In 2004, the study of exact 2D solids became a renewed focus,
as graphene was isolated from graphite,~\cite{Novoselov2004a} and its many 
fascinating properties were demonstrated. Graphite can be viewed
as a stack of graphene sheets held together by the weak van
der Waals bonding.

Graphene has a hexagonal arrangement of carbon atoms in
two dimensions, with a very strong covalent bonding between
them. One consequence is the extremely high mechanical
strength of the graphene sheet. Another key consequence is
that the electrons behave as if they are massless or relativistic
electrons (so-called Dirac electrons), whereas electrons in
standard semiconductors such as silicon behave as traditional
electrons with mass; this has been shown to lead to electron
mobilities a hundred times larger for graphene~\cite{Bolotin2008a} with the
potential for much faster electronics than with conventional
silicon (see the December 2012 special issue of MRS Bulletin
on graphene). Furthermore, graphene is neither a semiconductor 
nor a metal, rather it is exactly in between and can be called
a semimetal; technically, it is said to not have an electronic
energy gap. The ease of making graphene by mechanical
exfoliation (simply by peeling off graphene sheets from
graphite using Scotch tape) and the demonstration of various
properties of graphene led to Geim and Novoselov receiving
the Nobel Prize in Physics in 2010.

A simple question after the discovery of graphene would
be why not silicene (the silicon analog of graphene) or 
germanene (germanium analog) (i.e., the formation of 2D sheets
from other atoms from group IV of the periodic table).
Elements are grouped in the periodic table because of periodic
trends in their physical and chemical properties, and thus carbon, 
silicon, germanium, tin, and lead are expected to have
certain similar properties. A simple reason why the silicon
analog of graphene was not immediately considered is related
to the types of chemical bonding that are common for carbon
and silicon. Carbon is known to form various types of covalent
bonding (known as hybridization) involving two, three, and
four electrons forming such materials as ethylene, graphite,
and diamond, whereas silicon has been known to favor 
sharing four electrons equally, leading to bulk silicon. The key
to making 2D silicon, however, is to take a broader view of
two-dimensionality by not requiring all the atoms to form a
flat sheet; for chemists, this means one is not strictly looking
for the elusive $sp^2$ bonding for silicon.

\begin{figure*}[htp]
 \includegraphics[scale=0.75]{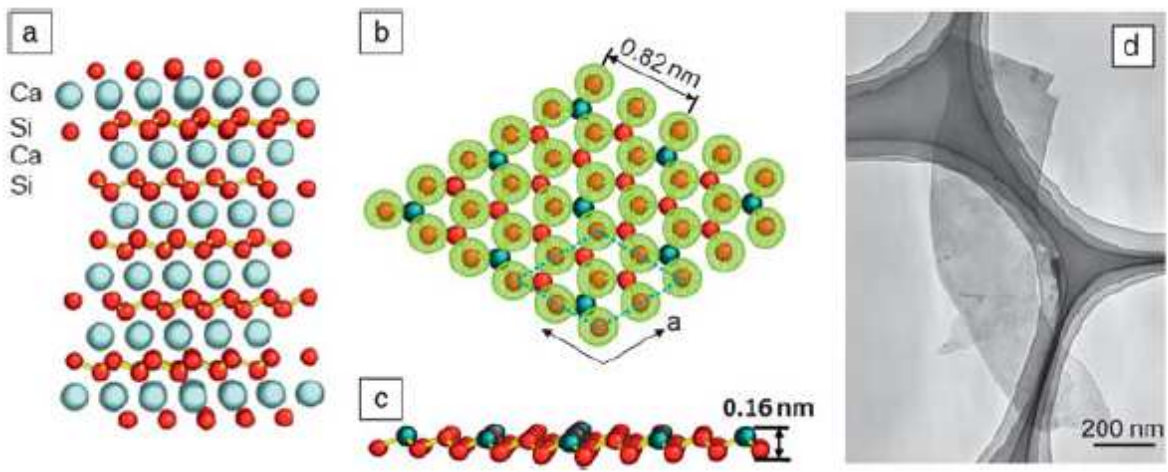}
 \caption{(a) Model of CaSi$_2$; (b-c) top and side view of Mg-doped silicon sheet capped
with oxygen; (d) TEM image of nanosheet. Reprinted with permission from Reference~[\onlinecite{Nakano2006a}].
\textcopyright 2006 Wiley-VCH Verlag GmbH \& Co. KGaA, Weinheim.}
\end{figure*}

This broader understanding was already achieved at least
as early as 1994 in a paper by Takeda and Shiraishi~\cite{Takeda1994a} in which
they asked what kind of 2D structures of silicon and germanium
might be possible. They carried out quantum mechanical ab
initio calculations that represented the state of the art in 
predicting the properties of materials. They found that structures
with minimum energies could be obtained if the two atoms in a
unit cell (the smallest repeat unit needed to generate the whole
infinite structure) are not in the same plane, leading to a sheet
that has “dimples” rather than both atoms in the same plane as
for graphene. In the process, they also obtained electronic
energy curves revealing the semi-metallic property, but they
did not know to look for the linear curves (so-called Dirac
cones) indicative of the Dirac electrons. There was also some
early work on making silicon nanosheets by chemical exfoliation 
and, in some cases, monolayers were reported, though
they were also functionalized and in most cases not very
stable.~\cite{Nakano2005a, Nakano2006a}

The demonstration of the presence of Dirac cones in
silicene first occurred in a paper published in 2007 by the
authors of this article.~\cite{Guzman-Verri2007a} While previous papers had used a
numerical method for computing the electronic energies, this
study used an analytical model that was able to prove, independent 
of whether the 2D sheet is flat or dimpled, Dirac
cones will result. This is also believed to be the first instance
of the use of the word “silicene” to describe the silicon 
analog of graphene.

All of this previous work remained fairly obscure until
$2009-2010$, when a French group provided the first hints
of the possible fabrication of silicene nanoribbons on a silver
substrate.~\cite{Kara2009a} Activity since has grown exponentially, and 
silicene is now recognized as a promising 2D material beyond
graphene.

\section{Growth of silicene}
Since a graphite-like form of silicene does not exist and all the
covalent bonds in bulk Si are equally strong, it is unlikely that
a mechanical exfoliation technique could be developed for
silicene. Hence, other techniques are needed. Two have been
reported to date.

\begin{figure}
  \includegraphics[scale=0.5]{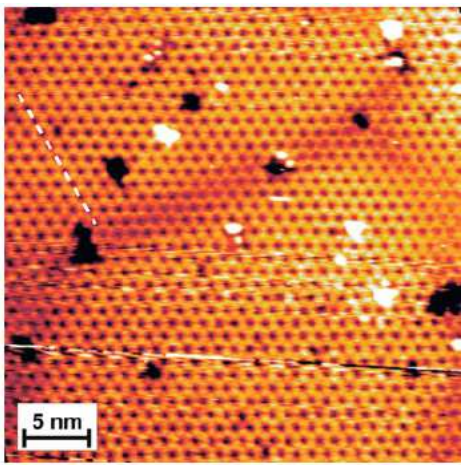}
  \caption{Scanning tunneling microscopic image of silicene on
Ag (111). Reprinted with permission from Reference~[\onlinecite{Vogt2012a}]. \textcopyright 2012
American Physical Society.} 
\end{figure}

\subsection{Wet chemistry method}

The first attempt at making an atomic layer of
silicon was via chemical exfoliation of calcium
disilicide~\cite{Nakano2005a, Nakano2006a} (Figure 1). Chemical exfoliation is
an attractive fabrication technique for nanostructures since it is a relatively inexpensive
and low-technology technique. The basic
idea is to use foreign atoms or ions to assist
in splitting bonds (primarily via steric and 
kinetic attacks), leading to free layers. This,
therefore, requires that the starting material
have a layered structure (i.e., certain weaker
bonds that can be preferentially broken while
preserving other bonds within the layers). Thus,
this technique is very effective for making 
graphene from graphite due to the weak interlayer
van der Waals bonds. One common side effect
of chemical exfoliation is a resulting functionalization of
the prepared materials, particularly from the foreign atoms.
However, if functionalization is desired, then this is an added
advantage.

A number of possible candidates exist for making silicene
via chemical exfoliation. Thus, calcium disilicide (CaSi$_2$) has
been used to prepare siloxene (a flat form of silicon with
attached OH groups), which can then be exfoliated to produce
siloxene nanosheets,~\cite{Nakano2005a} a form of functionalized silicene.
Mg-doped silicene sheets capped with oxygen were obtained
when Mg-doped CaSi$_2$ was directly exfoliated using propylamine 
hydrochloride; Mg doping was used because the doping
reduced the charges on the Ca and Si and, therefore, aided in
the exfoliation. On the other hand, polysilane could be 
exfoliated in an organic solvent to give silicon nanosheets covered
with organic groups instead of oxygen.~\cite{Okamoto2010a, Sugiyama2010a}

\begin{figure*}[htp]
 \includegraphics[scale=0.5]{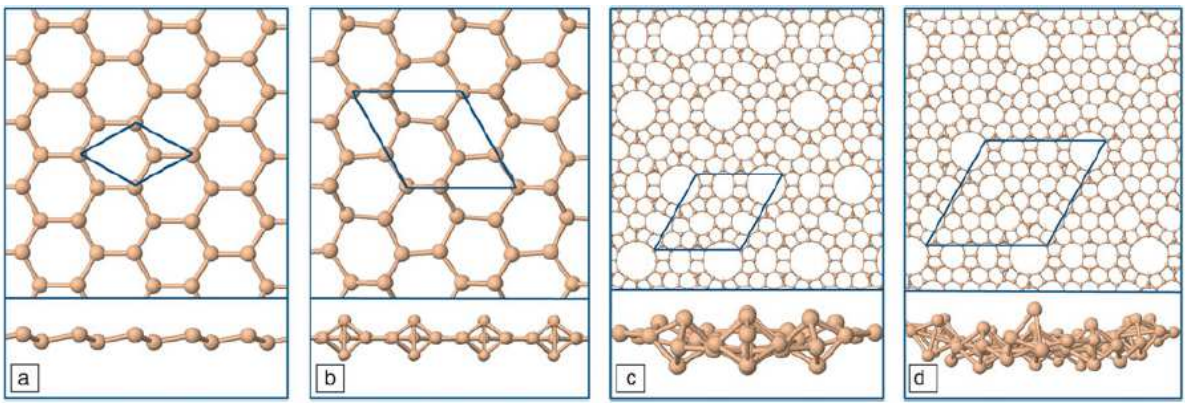}
 \caption{(a-d) Structures of a freestanding monoatomic layer of silicon. Each figure represents a minimum energy configuration found,
with the top drawing being a top view and the bottom drawing being a side view of the structure. Reproduced, in part, from Reference~[\onlinecite{Kaltsas2013a}].
 \textcopyright 2013 PCCP Owner Societies.}
\end{figure*}

\subsection{UHV deposition}
The first breakthrough in experimental silicene research was
the report of the fabrication of silicene nanoribbons on a 
silver substrate oriented in the ($110$) direction. The technique
used was an ultrahigh vacuum (UHV) deposition of silicon
atoms onto a metallic substrate. Deposition is one approach
to prevent 3D island formation, as bulk Si is more stable
than silicene by $1.56\,$ eV/atom.~\cite{Zhang2012a} Currently, successful 
fabrication of silicene sheets has been reported using an Ag ($111$)
substrate held at $150-300\,^\circ$C and a slow deposition rate of Si
below $0.1$ monolayer per minute~\cite{Vogt2012a, Lin2012a, Feng2012a, Chiappe2012a} (Figure 2). 
Silver has turned out to be an ideal substrate because of the low 
reactivity between Si and Ag and because of the compatibility of the
crystal structures and lattice constants.

There are still many unanswered questions about the growth
of silicene on Ag. Thus, different superstructures of silicene on
Ag have been reported, whether at one substrate temperature
or at different temperatures,~\cite{Lin2012a, Jamgotchian2012a} and there is not yet a 
conclusive explanation for all of these results. There appears to be a
correlation with substrate temperature;~\cite{Jamgotchian2012a} however, theoretical
modeling of structures (e.g., using molecular dynamics) is not
at the level of predicting the superstructures. Two other results
still under dispute concern the level of interaction between Si
and Ag electronic states, and the presence or not of the Dirac
electrons in silicene. These are related questions since they
both concern what happens to the electrons in silicene in the
presence of the Ag substrate; we will discuss this further
later in the article. Silicene has also been reported to have
been grown on ZrB$_2$~\cite{Fleurence2012a} and Ir ($111$).~\cite{Meng2013a}

\section{Properties of silicene}

\subsection{Structure}

Freestanding silicene can be viewed as a 2D material.
However, it is not completely flat, as is graphene. The most
commonly reported structure has been with alternate Si
atoms lying in planes that are separated from each other in the
direction perpendicular to the planes by $0.45\,$\AA (Figure 3a).

Nevertheless, it makes sense to still refer to these planes as
forming a buckled 2D sheet. In fact, such a sheet is very similar
to the ($111$) plane of Si.

A hypothetical flat silicene has been shown to be 
metastable via calculation of the phonon modes.~\cite{Cahangirov2009a} The buckled form
is also lower in energy by $30\,$ meV/atom~\cite{Durgun2005a} and has a binding
energy of $4.9\,$eV/atom, which is about $0.6\,$eV/atom lower than
for bulk Si.~\cite{Cahangirov2009a} More recent work now indicates that the 
honeycomb structure of Si might not be the lowest energy one;~\cite{Kaltsas2013a}
rather, other slightly more stable structures were obtained by
allowing the Si ($111$) plane to reconstruct. In fact, the other
structures, the so-called $ \sqrt{3} \times \sqrt{3}, 5 \times 5, $ and $ 7 \times 7$, (where the
expressions label the geometrical commensurability with the
basic unit cell) were found to be more stable than the silicene
structure (Figure 3b-d). These examples were chosen because
they are present in the surface reconstruction of bulk Si.
Nevertheless, it is expected that the structure that is obtained
in any given growth would be related to the growth conditions
and substrate.

Thus, the structure of silicene on Ag is much more complex
than previously thought, and a variety of superstructures have
been reported from both experiments and theory.~\cite{Vogt2012a, Feng2012a, Chiappe2012a, Lalmi2010a, Enriquez2012a, Kaltsas2012a, Gao2012a, Arafune2013a, Huang2013a} 
Growth conditions, particularly substrate temperature,~\cite{Huang2013a}
determine which superstructure is obtained. Some ambiguity
remains due in part to experimental characterization 
(e.g., scanning tunneling micrfoscopy) not associating a unique and 
well-defined structure to a given image.

\subsection{Elastic properties}

Two-dimensional materials require a new definition of their
elastic constants (e.g., the 3D bulk modulus is defined in terms
of isotropic pressure and volume change). One could define
in-plane stiffness,
\begin{align}
 C = h C_{11}\left[ 1 - \left(\frac{C_{11}}{ C_{12} }\right)^2 \right]
\end{align}
in terms of the 3D elastic constants $C_{11}$ and $C_{12}$, and an
effective thickness $h$.~\cite{Sahin2009a} One can also introduce 2D elastic
constants via~\cite{Wang2010a}
\begin{align}
 \gamma_{ij} = C_{ij} \times c_0 
\end{align}
where $c_0$ is the interlayer spacing of a hypothetical 3D 
supercell. Overall, it has been found that silicene is less rigid than
graphene. This is easily understood in terms of the buckled
structure of silicene compared to the flat structure of graphene.

With uniaxial strain, it is possible to open a bandgap, though
the dependence has been found to be nonlinear.~\cite{Zhao2012a} The largest
gap was found to be about $80\,$meV for a strain of about $8$\%.
Instability of the structure was obtained for a strain of $14-18$\%.
On the other hand, a biaxial tensile strain converted silicene
into a metallic state (for strain higher than
$7.5$\%) due to the lowering of the conduction
band at the $\Gamma$ point (the point in the Brillouin
zone with zero wave vector).~\cite{Liu2012a,Qin2012a} The latter
result is different from graphene, which remains
as a zero-gap material.

\subsection{Electronic properties}

The fascination with silicene is similar to that
for graphene - freestanding silicene has been
predicted to have Dirac cones just like graphene.
It should be repeated that this result, at first
sight, is not obvious due to the lower symmetry
of silicene compared to graphene. Nevertheless,
it has been shown rigorously that the degree
of buckling does not affect the existence of
the zero bandgap in the absence of spin-orbit
coupling, from symmetry arguments.~\cite{Guzman-Verri2007a} A similar
result derived from ab initio calculations is
given in Figure 4.

Figure 5 depicts the so-called band structure,
which is a plot of the electron energy versus
its wave vector (related to its momentum).
The point at the intersection of the Fermi level
(which basically represents the separation
between filled and unfilled electron states),
given here at zero energy, and the point K
shows the absence of the bandgap and the 
linear crossing of curves, the Dirac cone (when
the line is rotated). The band structure can
be recovered experimentally, with the most
direct technique currently in use being the
so-called ARPES (angle-resolved photoelectron spectroscopy) method. A photon of energy
$E$ is used to eject an electron of wave vector ${\bm k}$;
thus, simultaneous measurements of the two
quantities are possible.

\begin{figure}[htp]
 \includegraphics[scale=0.4]{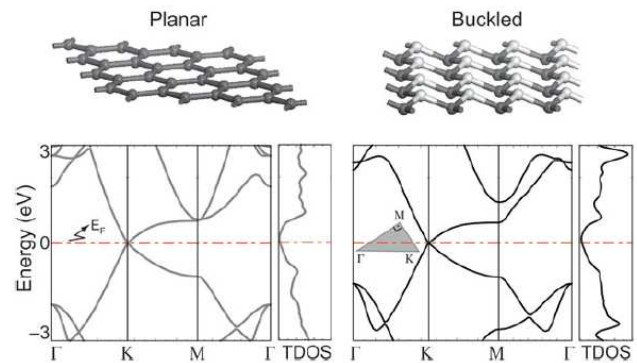}
 \caption{Band structure of flat and buckled silicene. Reprinted with permission from
Reference~[\onlinecite{Durgun2005a}]. \textcopyright 2005 American Physical Society. TDOS, total density of states.}
\end{figure}

\begin{figure}[h]
 \includegraphics[scale=0.4]{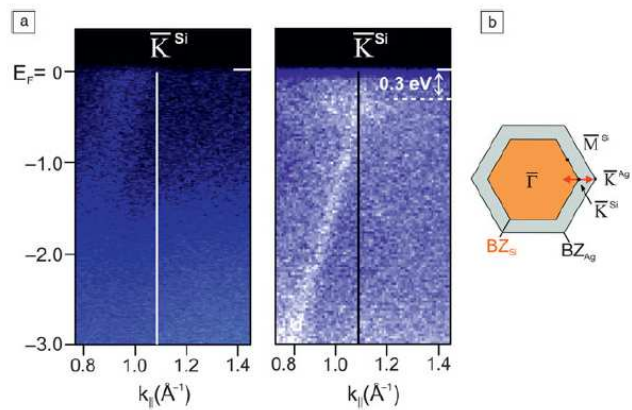}
\caption{ (a) Angle-resolved photoelectron spectroscopy of a clean silver surface (left)
and with silicene coverage (right); (b) Brillouin zone showing the location of various points.
Reprinted with permission from Reference~[\onlinecite{Vogt2012a}]. \textcopyright 2012 American Physical Society.}
\end{figure}

An example of an ARPES result is shown
in Figure 5, which appears fairly convincing in
showing a linear dispersion curve (the lighter
region in the second panel on the left figure).
Nevertheless, this result and another similar result~\cite{Chen2012a} have
recently been questioned.~\cite{Chen2013a, Guo2013a, Lin2013a, Wang2013a} 
Recent experiments have come
to the conclusion that the linear dispersion actually becomes
quadratic near the Dirac point.~\cite{Arafune2013a} Moreover, a search for massless Dirac electrons via the generation of the quantum Hall
effect turned out negative.~\cite{Lin2013a} The latter experimental results
have also received support from theoretical calculations,
revealing a strong hybridization between the Si and Ag electronic states~\cite{Lin2013a} and also others attaching the linear dispersion
to an $sp$ band origination from the silver. Indeed, the linearity
of the silicene electron is unlikely down to $3\,$eV, as implied in
Figure 5 (see the calculation of Wang and Cheng~\cite{Wang2013a}). Still, defense
of the Dirac cones persists, ranging from renewed identification of a linear tail to the silicene electronic dispersion even in the presence of a small gap~\cite{Avila2013a} to associating Dirac cones to
only certain superstructures.~\cite{Chen2013a} Clearly, the study of silicene on
silver remains complex and rich.

The electronic properties of silicene in the presence of a
vertical electric field (i.e., perpendicular to the plane) are
important for possible application as field-effect transistors. In
contrast to graphene, a bandgap opens for silicene due to the
lower symmetry of the buckled structure. In effect, the two
atoms in the unit cell feel different electric potentials since
they are at different heights. Two early calculations both obtained
a fairly linear dependence of the gap on the electric field;~\cite{Ni2012a, Drummond2012a}
however, they obtained rates that differ by a factor of two.

It was noted previously that the bandgap is zero in the
absence of spin–orbit coupling, which can split the bands and
lead to a small gap opening. There are two contributions to the
magnitude of the spin–orbit gap. One is the magnitude of the
atomic spin–orbit coupling; the other relates to symmetry 
reduction due to buckling. It turns out that both effects contribute to
a larger spin–orbit gap for silicene compared to graphene. It is,
in fact, negligible for graphene, while it is about $1.55\,$meV for
silicene.~\cite{Liu2011a} While still small, it is sufficient to help demonstrate
the quantum spin Hall effect; this effect arises from the existence of a bulk gapped state and gapless conducting edge states
at the boundaries, an example of a topological insulator. An
interesting proposal is to control the bandgap using an external
electric field,~\cite{Ezawa2012a, Ezawa2012b, Dyrdal2012a} transforming silicene from a 
topological insulator into a band insulator. Indeed, silicene has been predicted to
have an extremely rich phase diagram of topological states with
unique quantum states of matter such as a hybrid quantum Hall-quantum anomalous Hall state (the anomalous effect being the
well-known quantum Hall effect in the absence of an external
magnetic field) and a so-called valley-polarized metal (resulting
from electron transfer from a conduction valley to a different
hole valley), leading to the new field of spin valleytronics.~\cite{Ezawa2012c, Ezawa2013a, Tabert2013a}

\subsection{Optical}

In any study of silicon-based materials, optical properties are
an area of interest since one of the ``holy grails'' of materials
science is silicon-integrated photonics. Bulk Si is known to
be a poor light emitter due to its indirect bandgap. There have
been few studies on the optical properties of silicene to date
due in part to the zero gap for the freestanding sheet and to
the increased difficulty in computing optical properties using
density functional theory (DFT). The latter is known to be 
deficient in computing excited-state properties, necessary to describe
optical properties. Furthermore, the semi-metallic nature of
gapless silicene reduces the possible contribution due to 
electron screening, complicating the theory, particularly for single-particle 
ones. Experimentally, no results have yet been reported,
due to the difficulty in peeling silicene from the Ag substrate.

Early calculations on hydrogenated silicon nanosheets
(the effect of hydrogenation on the electronic properties is
covered in more detail in the next section) of different 
orientations have revealed the possibility of obtaining direct energy
gaps of the order of $2-3\,$eV.~\cite{Lu2009a} We note, however, that detailed
structures of the nanosheets were not provided, and they 
appear to be flat rather than buckled as we understand them to
be now. Furthermore, those calculations were done using the
generalized gradient approximation to DFT, which is known
to underestimate the bandgap. Calculations that fix the previous
problems~\cite{Pulci2012a} show that there can be a blueshift of the transition
by about $1\,$eV upon exciton formation and also an enhanced
fundamental oscillator strength, which is about $500$ times
stronger than for graphane (the hydrogenated form of graphene);
the latter has been attributed to the different nature of the 
lowest conduction states (being mostly localized on silicon atoms
for silicane (the hydrogenated form of silicene) and mostly on
hydrogen atoms for graphane).

One would expect the infrared properties to be very similar
to those of graphene since those are due to the zero gap, linear
energy dispersion, and the 2D nature. Indeed, the universality
of the optical absorption (proportional to the universal fine
structure constant) based upon a single-particle picture has
been predicted.~\cite{Bechstedt2012a}

Another aspect missing from the above is a study of excitons
(the collective behavior of electrons and holes in optically
excited materials) in silicene. It should be pointed out that
even the excitonic spectrum of graphene is not fully understood.
Recent calculations~\cite{Wei2013a} indicate that silicene also has resonant
excitons like graphene.

\subsection{Functionalization}

In spite of the interest in pure silicene, there are a number
of reasons why functionalization is of some importance. First
and foremost, growth processes might naturally lead to 
functionalized sheets (e.g., via wet chemical methods). Second,
the additional functionality introduced by adding various other
atoms enlarges the variety of properties attainable. Four 
general types of functionalizations have been studied for silicene:
hydrogenation, halogenation, by metals, and by organic groups.

Hydrogenation is, by far, the most studied.~\cite{Lu2009a, LewYanVoon2010a, Garcia2011a, Houssa2011a, Jose2011a, Ding2012a} Hydrogenation
of graphene was studied early on as a way of opening an
energy gap; hence, it is natural to consider it for silicene as well.
Also, experimentally, it is often possible to obtain hydrogenated
samples. Theoretically, incorporation of hydrogen atoms is
the standard approach to saturating dangling bonds. Finally,
from an application perspective, one can also envision 
hydrogen storage as an important goal of silicon (and other material)
research. The basic results are that full hydrogenation, with
the resulting material being called silicane, does lead to a
bandgap opening.~\cite{LewYanVoon2010a} Boat-like and chair-like configurations
are found to be the most stable ones but with very different
bandgaps ($2.9\,$ eV direct and $3.8\,$eV indirect, respectively,
using many-body perturbation theory~\cite{Houssa2011a}). More interestingly,
half-hydrogenation is found to lead to a direct-gap insulator
with a gap of $1.74 -  1.79\,$eV for the chair configuration and a
ferromagnetic state.~\cite{Wang2012a, Zhang2012b} The latter has been attributed to the
unpaired 3p electrons on the unhydrogenated silicon sites.
On the basis of mean-field approximation arguments, the Curie
temperature was estimated to be $122-300$ K.~\cite{Wang2012a, Zhang2012b} One approach to
obtaining half-hydrogenation is to first fully hydrogenate
silicene and then apply a vertical electric field.~\cite{Gang2013a}

Fluorination performs a similar function as hydrogenation.
It has been found that fluorinated silicene is more stable than
silicane and with a smaller bandgap.~\cite{Garcia2011a, Ding2012a} Bromine was found
to lead to even more interesting behavior.~\cite{Zheng2012a} Half-brominated
silicene is an antiferromagnetic half-metal, whereby one spin
channel is metallic, and the other is semiconducting with a
$1.73\,$eV energy gap.

Metal adatoms display various states and properties, as
one can expect from their variety. Even for graphene, metal
adatoms have led to a variety of behaviors such as 
superconductivity with lithium and catalytic properties with transition
metals. We refer the reader to the paper by Lin and Ni for an
extensive study.~\cite{Lin2012b} A general conclusion is that metal adatoms
bind much more strongly to silicene than to graphene. The
bonding is covalent in all cases except for alkali metal atoms.

\subsection{Other properties}
In addition to the previous extensive studies, we would like to
mention some other work on the properties of silicene. A few
papers have been published on various thermal properties.~\cite{Li2012a, Hu2013a}
In particular, the in-plane thermal conductivity of silicene is
found to be about an order of magnitude lower than for bulk Si.
Vacancy defects lead to further reduction due to phonon-defect
scattering. Furthermore, the thermal conductivity of silicene is
found to initially increase with tensile strain,~\cite{Hu2013a} opposite to what
is observed for graphene. The different behavior was traced to
the fact that all the vibrational modes for graphene softened with
tensile strain, whereas this is true for the longitudinal and 
transverse modes for silicene but not for the flexural modes 
(perpendicular to the layer). The latter competition between phonon
softening and stiffening for silicene leads to a nonmonotonic
behavior of the thermal conductivity as a function of strain.

\section{Silicene nanoribbon}
Graphene nanoribbons garnered interest as a means of 
opening a bandgap in graphene through the quantum confinement
effect. For silicene, it was also the case that the first reported
fabrication was actually of the nanoribbons rather than of the
sheet.~\cite{Vogt2012a} Indeed, from a growth point of view, nanoribbons were
very attractive due to the high uniformity of single 
nanoribbons (1.6 nm in width) and of arrays of the nanoribbons. The
nanoribbons were initially reported to be grown on Ag (110),
though this has also now been achieved on Au (110).~\cite{Tchalala2013a} They
were also observed to be much more stable to molecular 
oxygen that bulk Si.~\cite{Padova2011a}

The electronic properties of freestanding silicene nanoribbons
are fairly straightforward. Thus, armchair silicene 
nanoribbons (ASiNR) were all found to be nonmagnetic~\cite{Ding2009a} but could
be metals or semiconductors. The zigzag silicene nanoribbons
(ZSiNR) were found to have an antiferromagnetic 
semiconducting ground state. Here, armchair and zigzag describe
nanoribbons with two different shapes of edges. These results
are the same as for graphene nanoribbons.

Very large values of magnetoresistance have been predicted in
silicene nanoribbons using first-principle calculations.~\cite{Kang2012a} Silicene
nanoribbons with zigzag configurations have ferromagnetic
states at the edges. By attaching the ends of the nanoribbon to
electrodes and applying parallel magnetic fields to them, the
nanoribbon shows a parallel spin configuration along its length.
By applying antiparallel magnetic fields, the nanoribbon shows
an antiparallel spin configuration. Upon applying a bias field
between the electrodes, a current flows from one electrode to the
other, and it increases with the bias field in an approximate 
linear fashion for both the parallel and antiparallel configurations.
The calculated current of the parallel configuration is, however,
several orders of magnitude greater than that of the 
antiparallel configuration as a result of distinct selection rules between
electronic states. The percent change in the resistance is huge
($10^6\,$\%), and it is comparable to that of graphene nanoribbons.~\cite{Kim2008a}

\section{Applications}
Given the similarity of silicene to graphene, many of the same
potential applications of graphene have been considered for
silicene. Experimentally, one is still far from any device
fabrication given that silicene growth is still in its infancy and
much about its properties remains to be characterized. Hence,
we only provide a very brief survey of some of the 
applications being envisioned for silicene.

The most obvious such potential application is in 
nanoelectronics as transistors. Graphene transistors have already been
demonstrated. The advantages of silicene are that it is 
compatible with current silicon nanoelectronics, and one can more
easily open a gap, whether through coupling with a substrate,
strain, or simply by using a vertical electric field.~\cite{Ni2012a, Drummond2012a} The latter
is obviously attractive for field-effect transistors (FETs). One
current obstacle is obtaining silicene on an insulator, though
this is being actively pursued.~\cite{Houssa2010a, Liu2013a} Notwithstanding, the per-
formance of a silicene nanoribbon FET has been modeled.~\cite{Li2012a}
It is found that short ASiNR FETs have large current on/off
ratios of over $10^6$, and the output characteristic exhibits a 
saturation current, an effect absent for graphene nanoribbons.

Silicene has been proposed as an ideal spintronics and
valleytronics material. The spin–orbit interaction for silicene is
predicted to be relatively large compared to that for graphene,
simply because the atomic spin–orbit coupling is larger for
silicon than for carbon.~\cite{Liu2011a} This opens up the possibility of spin
and valley physics otherwise difficult to observe in graphene.~\cite{Tahir2013a}
It has been predicted that an electric current of a definite spin
and valley label could be isolated.~\cite{Ezawa2013a, Tabert2013a} Since the valley struc-
ture can be controlled by an electric field, this could lead to an
electric-field controlled spin filter.~\cite{Wang2012a, Tsai2013a}.

Recently, it has been suggested that silicene may be suitable for energy storage applications.~\cite{Tritsaris2013a} The energy density of
Li-ion batteries depends on the specific charge capacity of the
electrodes. Being of atomic thickness, silicene could serve as a
high-capacity host for Li. First-principles calculations find that
freestanding single-layer and double-layer silicene have 
binding energies of about $2.2\,$eV per Li atom, which do not vary
much with respect to Li content and have smaller diffusion 
barriers ($\leq 0.6\,$eV) than those of bulk silicon and silicon nanowires.
Binding energies of silicene with other alkali, alkali-earth metals,
groups III and IV metals, and transition metals have been 
calculated and were found to be stronger than those with graphene.~\cite{Lin2012a}
Electronic band structures of lithiated silicene have been studied
by first-principles calculations and showed that upon complete
lithiation, the band structure of silicene transformed from a 
zero-gap semiconductor to a $0.368\,$eV bandgap semiconductor.~\cite{Osborn2012a}

\section{Summary}
Silicene has been touted as the next graphene since it is 
predicted to display the same linear electronic dispersion, yet one
can more easily induce a bandgap opening due to the reduced
crystal symmetry. Additionally, it is compatible with current
silicon microelectronics. While its fabrication has now been
reproducibly verified, this has so far only been achieved on
metallic substrates, limiting its potential applications in 
field-effect transistors. Promising avenues for both theoretical
and experimental work also include ways in which it differs
from graphene. One such activity that has been researched
theoretically is in spintronics, whereby both the much larger
spin–orbit coupling of silicon as compared to diamond and
the symmetry difference between silicene and graphene have
been exploited to propose new physics and applications.
An explosive growth in silicene research will likely depend on
the materials community achieving a breakthrough in making
silicene easily, abundantly, and on a variety of substrates.

\section{Acknowledgments}
Our early research on silicene was partially funded by the
National Science Foundation. Writing of this article was
facilitated by funds from The Citadel Foundation, the Traubert
Endowed Funds, and the US Department of Energy, Office of
Basic Energy Sciences under contract no. DE-AC02-06CH11357.

\end{document}